\documentclass[12pt,preprint]{aastex}

\usepackage{amsmath}
\usepackage{hyperref}
\usepackage[flushleft]{threeparttable}

\hypersetup{
    colorlinks=true,
    linkcolor=blue,
    filecolor=blue,      
    urlcolor=blue,
    citecolor=blue,
}

\shorttitle{Equilibrium Temperature of Planets}
\shortauthors{M\'{e}ndez \& Rivera-Valent\'{i}n}
\slugcomment{ M\'{e}ndez, A. \& Rivera-Valent\'{i}n, E. G. (2017) \emph{ApJL}, 837, L1}

\begin{document}


\title{The Equilibrium Temperature of Planets in Elliptical Orbits}


\author{Abel M\'{e}ndez\altaffilmark{1}}
\affil{Planetary Habitability Laboratory, University of Puerto Rico at Arecibo\\PO Box 4010, Arecibo, PR 00614}
\email{abel.mendez@upr.edu}

\author{Edgard G. Rivera-Valent\'{i}n}
\affil{Arecibo Observatory, Universities Space Research Association\\ HC 3 Box 53995, Arecibo, PR 00612}
\email{ervalentin@usra.edu}

\altaffiltext{1}{Corresponding Author}


\begin{abstract}

There exists a positive correlation between orbital eccentricity and the average stellar flux that planets receive from their parent star. Often, though, it is assumed that the average equilibrium temperature would correspondingly increase with eccentricity. Here we test this assumption by calculating and comparing analytic solutions for both the spatial and temporal averages of orbital distance, stellar flux, and equilibrium temperature. Our solutions show that the average equilibrium temperature of a planet, with a constant albedo, slowly decreases with eccentricity until converging to a value $90\%$ that of a circular orbit. This might be the case for many types of planets (\emph{e.g.}, hot-jupiters); however, the actual equilibrium and surface temperature of planets also depend on orbital variations of albedo and greenhouse. Our results also have implications in understanding the climate, habitability and the occurrence of potential Earth-like planets. For instance, it helps explain why the limits of the habitable zone for planets in highly elliptical orbits are wider than expected from the mean flux approximation, as shown by climate models.

\end{abstract}

\keywords{stars: planetary systems, planets and satellites: fundamental parameters, planetary habitability, equilibrium temperature, habitable zone, eccentricity.}


\section{Introduction} \label{sec:intro}


The climate and potential habitability of planets depend on a complex interaction of many stellar and planetary properties \citep{2011AsBio..11.1041S}. For example, Earth experiences a small annual change of 4 K in its average global surface temperature\footnote{Berkeley Earth: \href{http://berkeleyearth.org/data/}{http://berkeleyearth.org/data/}} mostly due to its obliquity and distribution of land and ocean areas and not its low eccentricity (\emph{i.e.,} $e>0.0167$); however, high eccentricities may have a strong effect on the climate of exoplanets or exomoons (\emph{e.g.,} stability of liquid water between seasonal extremes). This is a very important effect to consider since 75\% of all exoplanets with known orbits have orbital eccentricities larger than Earth\footnote{NASA Exoplanet Archive: \href{http://exoplanetarchive.ipac.caltech.edu/}{http://exoplanetarchive.ipac.caltech.edu/}}.


The effects of orbital eccentricity on planetary habitability is a complex problem that has been explored by one- and three-dimensional climate models \citep[\emph{e.g.},][]{2002IJAsB...1...61W, 2010ApJ...721.1295D, 2013Icar..226..760D, 2014ApJ...791L..12W, 2014AsBio..14..277A, 2015Icar..250..395D, 2015P&SS..105...43L, 2016A&A...591A.106B, 2017ApJ...835L...1W}. The suitability for planets to maintain surface liquid water can be constrained by their orbital location as defined by the habitable zone (HZ), the circular region around a star in which liquid water could exist on the surface of a rocky planet \citep{1959PASP...71..421H, 1964hpfm.book.....D, 1993Icar..101..108K}. Estimates on the occurrence of Earth-like worlds around stars $\eta_{\oplus}$ depend on recognizing those in the HZ \citep{2015IJAsB..14..359T}.

The conservative inner edge of the HZ is given by the runaway of liquid water (\emph{i.e.,} surface temperatures above $\sim$340 K), corresponding to a rapid increase in stratospheric water vapor content, which leads to loss of an Earth-like ocean inventory within four to five billion years. The conservative outer edge is given by the maximum greenhouse necessary to keep temperatures above freezing (\emph{i.e.,} surface temperature above 273 K). These limits assume an Earth-like planet with N$_{\text{2}}$-CO$_{\text{2}}$-H$_{\text{2}}$O atmospheres and supported by a carbonate-silicate cycle \citep{1993Icar..101..108K}. Further developments on the limits of the HZ for main-sequence stars are discussed elsewhere \citep[\emph{e.g.},][]{2011ApJ...734L..13P, 2012AsBio..12....3J, 2013ApJ...771L..45Y, 2013AsBio..13..715S, 2013ApJ...765..131K, 2014ApJ...787L..29K, 2016ApJ...819...84K, 2016ApJ...817L..18K,2017arXiv170107513K}. The HZ for circular orbits can be defined either in terms of orbital distance, stellar flux, or equilibrium temperature. Usually, a planetary albedo of 0.3 (similar to Earth or Jupiter) is used for the calculation of the equilibrium temperature as a basis for comparisons.


The equilibrium temperature is useful for comparing the temperature regimes of planets, but is not sufficient to draw conclusions on their climate \citep{2013A&A...554A..69L}. Surface temperatures are controlled by a complex interaction of albedo and greenhouse effects. Planetary albedo can be estimated from equilibrium temperatures determined spectroscopically or from planet/star flux ratios \citep{2011ApJ...729...54C, 2012ApJ...757...80C}. In practice, it might be easier to go the other way around and estimate equilibrium temperatures from planetary albedo using thermal and optical phase curve photometry, which requires less light gathering power than spectroscopy \citep{2009Icar..204...11M, 2016A&A...587A.149V}.


The equilibrium temperature $T_{eq}$ of the planet results from a balance between the incident stellar flux $F_{\star}$ from the host star and that absorbed by the planet \citep{2007A&A...476.1373S}, where $F_{\star}$ and $T_{eq}$ are given from the Stefan-Boltzmann law
\begin{equation} \label{eq:Fstar}
F_{\star} = \frac{L_{\star}}{4 \pi d^2} = \frac{\sigma R_{\star}^2 T_{\star}^4}{d^2},
\end{equation}
\begin{equation} \label{eq:Teqstar}
T_{eq}= \left[ \frac{(1-A) F_{\star}}{4 \beta \epsilon \sigma} \right]^ \frac{1}{4} = \left[ \frac{(1-A) L_{\star}}{16 \beta \epsilon \sigma \pi d^2} \right]^ \frac{1}{4} = T_{\star} \sqrt{\frac{R_{\star}}{2d}} \left( \frac{1-A}{\beta \epsilon} \right) ^\frac{1}{4},
\end{equation}
where $L_{\star}$, $R_{\star}$, $T_{\star}$ are the luminosity, radius, and the effective temperature of the star, respectively, $d$ is the distance of the planet from the star, $\epsilon$ is the broadband thermal emissivity (usually $\epsilon \approx 1$), and $\sigma$ is the Stefan-Boltzmann constant. The planetary albedo or bond albedo $A$ is the fraction of incident radiation, over all wavelengths, which is scattered by the combined effect of the surface and atmosphere of the planet. The factor $\beta$ is the fraction of the planet surface that re-radiates the absorbed flux, $\beta = 1$ for fast rotators and $\beta \approx 0.5$ for tidally locked planets without oceans or atmospheres \citep{2011ApJ...736L..25K}. Equation \ref{eq:Teqstar} ignores other energy sources such as formation energy, tidal deformation, and radiogenic decay. For simplicity, these equations can be expressed using solar and terrestrial units where the distance between the planet and the star, $r=d/a_{\earth}$, are in astronomical units (AU), and the stellar flux, $L=L_{\star}/L_{\sun}$, in solar units such that $F$ and $T_{eq}$ become
\begin{equation} \label{eq:F}
F = \frac{L}{r^2} ,
\end{equation}
\begin{equation} \label{eq:T}
T_{eq}=T_o \left[ \frac{(1-A)L}{\beta \epsilon r^2} \right] ^ \frac{1}{4},
\end{equation}
where $T_o = [L_{\sun}/(16\pi\sigma a_{\earth}^2)]^\frac{1}{4}$ = 278.5 K (\emph{i.e.}, the equilibrium temperature of Earth for zero albedo). The derivations presented in this paper use solar and terrestrial units from now on as shown in equations \ref{eq:F} and \ref{eq:T}.


Surface temperatures, though, depend on the equilibrium temperature and greenhouse effect of the planets. The normalized greenhouse effect $g$ can be used to connect the equilibrium and surface temperatures as
\begin{equation} \label{eq:g}
g = \frac{G}{\sigma T_s^4} = 1 - \left( \frac{T_{eq}}{T_s} \right)^4,
\end{equation}
where $G$ is the greenhouse effect or forcing (W/m$\textsuperscript{2}$), and $T_s$ is the surface temperature \citep{1989Nat..342..758}. The normalized greenhouse is very convenient since it summarizes a complex planetary property in a unitless quantity, as the planetary albedo does. Together equations \ref{eq:T} and \ref{eq:g} combine to give the surface temperature of a planet as
\begin{equation} \label{eq:Tsurf}
T_s = T_o \left[ \frac{(1-A) L}{\beta \epsilon (1-g) r^2} \right] ^\frac{1}{4}.
\end{equation} The final surface temperature of a planet is controlled by its heat distribution $\beta$, emissivity $\epsilon$, bond albedo $A$, and greenhouse $g$; all properties that depend on combined surface and atmospheric properties. It is also important to clarify that the average of equation \ref{eq:Tsurf} is in fact the effective surface temperature $\langle T_{se} \rangle$ and not necessarily the actual surface temperature $\langle T_{s} \rangle$, depending on the spatial and temporal scales considered (\emph{i.e.}, global, daily, or annual averages). For example, the annual global average surface temperature of Mars is $\langle T_s \rangle \approx$ 202 K, but its effective surface temperature is $\langle T_{se} \rangle \approx$ 214 K \citep{2013Icar..223..619H}. This is an important distinction necessary to calculate greenhouse effects.


Equations \ref{eq:F} and \ref{eq:T} are more complicated for elliptical orbits. It is often assumed that the average equilibrium temperature $\langle T_{eq} \rangle$ of a planet in an elliptic orbit can be simply calculated from its average stellar flux $\langle F \rangle $ or distance $\langle r \rangle $. \cite{2013A&A...554A..69L} recognized that the temporal average of the equilibrium temperature is lower than the equilibrium temperature computed from the temporally averaged flux, although they did not calculate it. Thus, the expressions
\begin{equation} \label{eq:Twrong1}
\langle T_{eq} \rangle \neq T_o \left [ \frac{(1-A) \langle F \rangle}{\beta \epsilon} \right ] ^ \frac{1}{4}, 
\end{equation}
\begin{equation} \label{eq:Twrong2}
\langle T_{eq} \rangle \neq T_o\left[ \frac{(1-A)L}{\beta \epsilon \langle r \rangle^2}\right] ^\frac{1}{4}, 
\end{equation}
are incorrect interpretations that have been used extensively in the literature. For example, \cite{2011ApJ...736L..25K} (equation 6 in their paper) and \cite{2011ApJ...726...82C} (equation 9 in their paper) suggested a formulation equivalent to equation \ref{eq:Twrong1}. Also \cite{2013ApJ...774...27H} (table 2 in their paper) used this interpretation to calculate the equilibrium temperature for planets in elliptical orbits. Formulations similar to equations \ref{eq:Twrong1} or \ref{eq:Twrong2} might introduce large errors in estimates of the average equilibrium temperature of planets, especially for large eccentricities. 


In this paper, we computed analytical solutions for the average orbital distance, stellar flux, and equilibrium temperature of planets in elliptical orbits. Orbital averages were computed with respect to spatial (\emph{i.e.}, geometric interpretation) and temporal (\emph{i.e.}, physical interpretation) coordinates for comparison purposes. In \S\ref{sec:space} and \S\ref{sec:time}, we show correct formulations using both spatial and temporal averages for elliptical orbits, respectively. To our knowledge, we determined a new analytical solution for the average equilibrium temperature of planets in elliptical orbits. In \S\ref{sec:hz}, we apply our result to define an effective orbit for the HZ. The implications of our results are discussed in \S\ref{sec:discussion}. Our study is motivated by planets in the HZ, but it is applicable to any planet in elliptical orbit with nearly constant albedo (\emph{e.g.}, Hot-Jupiters).


\section{Spatial Averages for Elliptical Orbits} \label{sec:space}

\begin{figure}
\epsscale{1.0}
\plotone{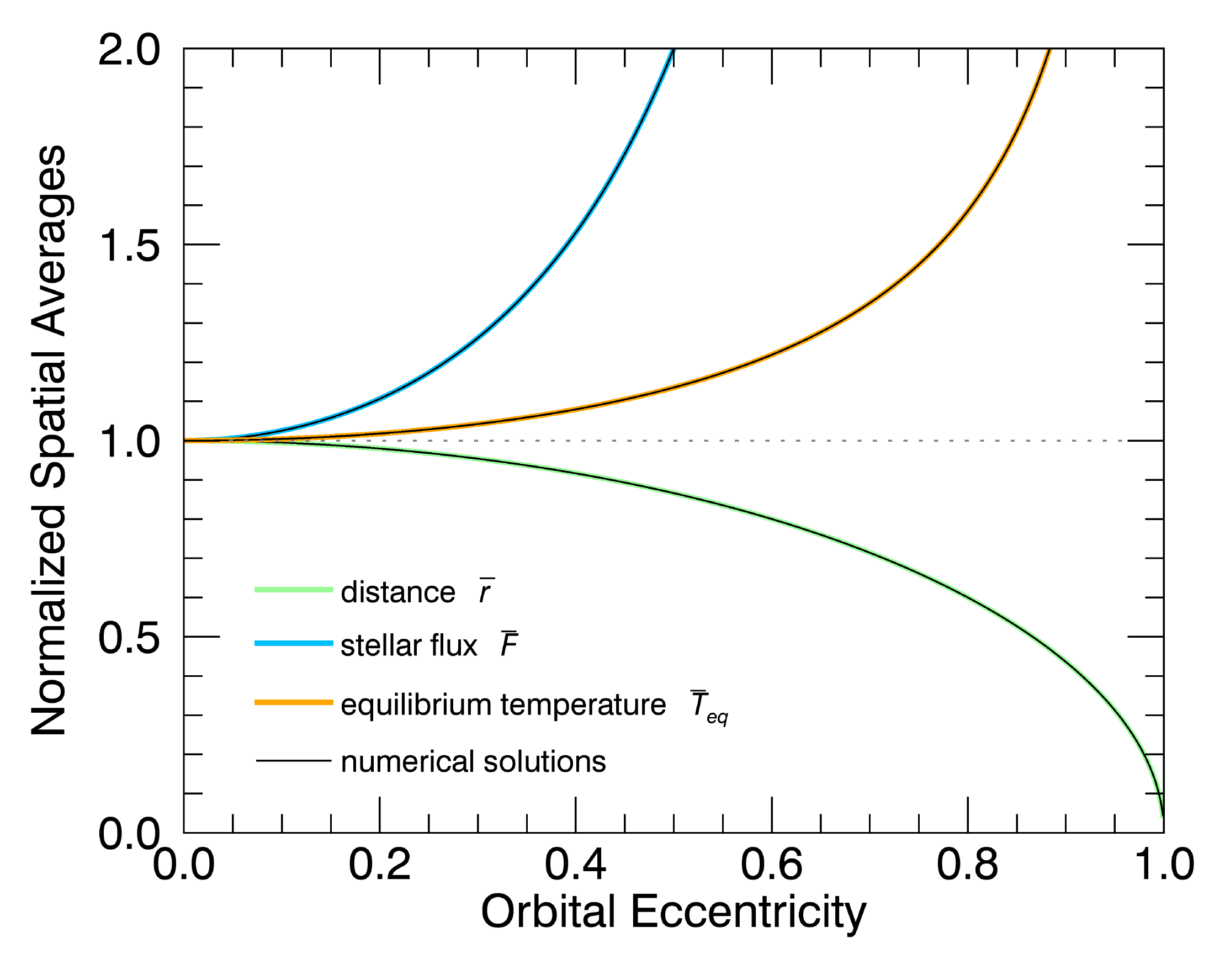}
\caption{Comparison of spatial averages of orbital distance, stellar flux, and equilibrium temperature as a function of eccentricity (normalized with respect to the circular orbits values). Both analytic (color lines) and numerical solutions (black lines within the color lines) are shown.}
\label{fig.s}
\end{figure}

Spatial averages for orbital distance $\overline{r}$, stellar flux $\overline{F}$, and equilibrium temperature $\overline{T}_{eq}$ do not have a practical observational application, but were computed here for comparison purposes. They could be calculated with respect to the eccentric anomaly $E$ or true anomaly $f$ (orbital longitude). Equations \ref{eq:F} and \ref{eq:T} were integrated over the true anomaly (making the substitution $r=a(1-e^2)/(1+e\cos{f})$) for a full orbit to obtain the corresponding averages,
\begin{equation} \label{eq:rs}
\overline{r} = \frac{1}{2\pi} \int_0^{2\pi} \! r \, \mathrm{d}f = a\sqrt{1-e^2},
\end{equation}
\begin{equation} \label{eq:Fs}
\overline{F} = \frac{1}{2\pi} \int_0^{2\pi} \! F \, \mathrm{d}f = \frac{L(2+e^2)}{2a^2(1-e^2)^2},
\end{equation}
\begin{align} \label{eq:Ts}
\overline{T}_{eq} = \frac{1}{2\pi} \int_0^{2\pi} \! T_{eq} \, \mathrm{d}f &= T_o\left[ \frac{(1-A)L}{\beta \epsilon a^2}\right] ^ \frac{1}{4} \frac{2}{\pi \sqrt{1-e}} \; \mathbf{E} \left ( \sqrt{\frac{2e}{1+e}} \right ) \\
&\approx T_o\left[ \frac{(1-A)L}{\beta \epsilon a^2}\right] ^\frac{1}{4} \left[ 1 + \tfrac{7}{16}e^2 + \tfrac{337}{1024}e^4 + O(e^6) \right],
\end{align}
where $\mathbf{E}$ is the complete elliptic integral of the second kind \citep{MathWorld, GSL} and $a$ is the semi-major axis. Equation \ref{eq:rs} is sometimes used as the average distance of planets in elliptical orbits. This is a geometric average and the more physically meaningful quantity, the temporal average (see \S\ref{sec:time}), should be used instead. The equations for stellar flux (equation \ref{eq:Fs}) and equilibrium temperature (equation \ref{eq:Ts}) are new results of this paper. Both increase with eccentricity while the average distance decreases (figure \ref{fig.s}); however, these results are of little practical value since they are geometric averages.


\section{Temporal Averages for Elliptical Orbits} \label{sec:time}

\begin{figure}
\epsscale{1.0}
\plotone{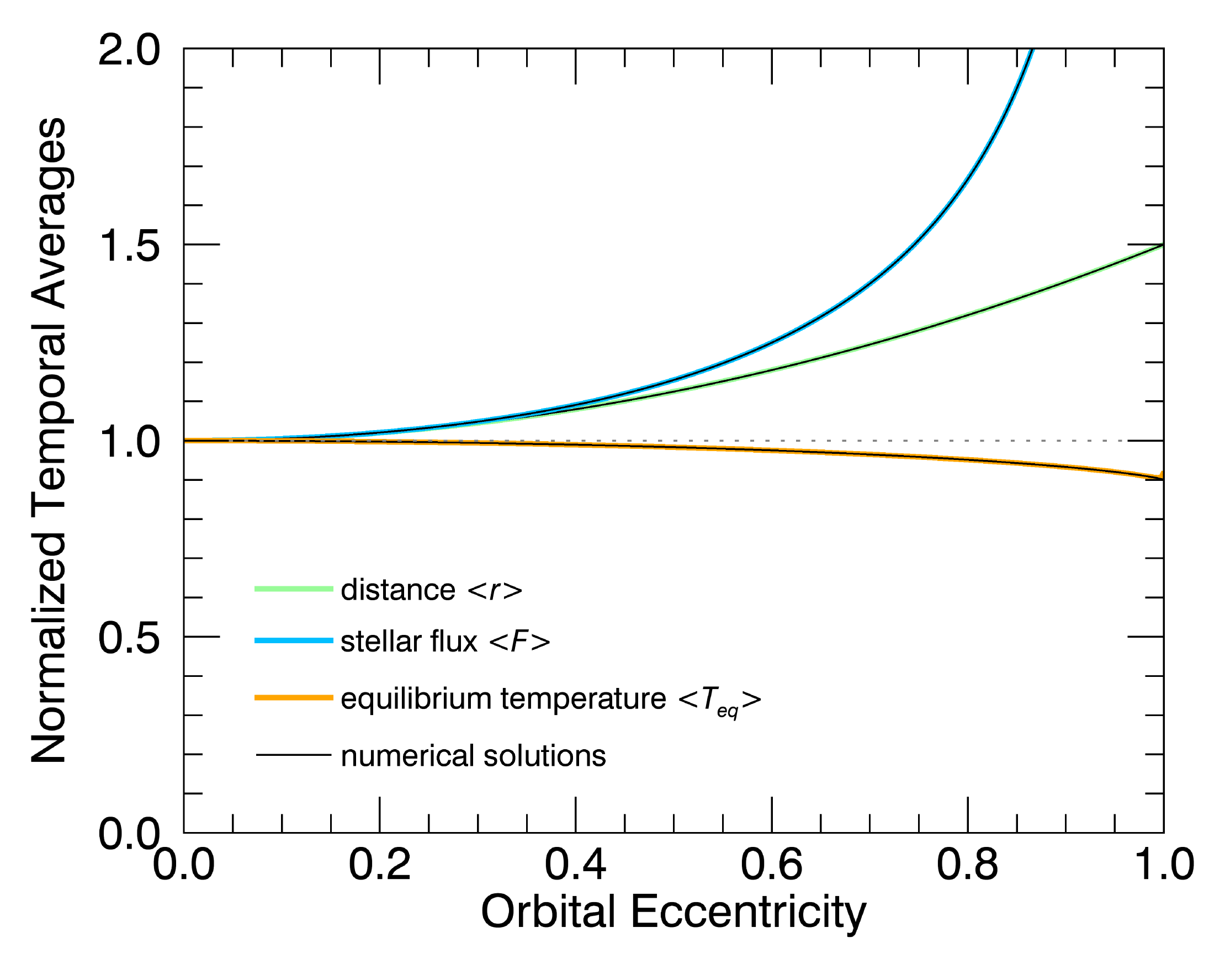}
\caption{Comparison of temporal averages of orbital distance, stellar flux, and equilibrium temperature as a function of eccentricity (normalized with respect to the circular orbits values). Both analytic (color lines) and numerical solutions (black lines within the color lines) are shown.}
\label{fig.t}
\end{figure}

Temporal averages for orbital distance $\langle r \rangle$, stellar flux $\langle F \rangle$, and equilibrium temperature $\langle T_{eq} \rangle$ were computed. They could be calculated with respect to time $t$ or the mean anomaly $M$. Equations \ref{eq:F} and \ref{eq:T} were integrated over time (making the substitutions $r=a(1-e\cos{E})$, $M=E-e\sin{E}$, and $M=(2\pi/T)t$) for a full orbital period $T$ to obtain the corresponding averages
\begin{equation} \label{eq:rt}
\langle r \rangle = \frac{1}{T} \int_0^T \! r \, \mathrm{d}t = a\left ( 1+\frac{e^2}{2} \right ),
\end{equation}
\begin{equation} \label{eq:Ft}
\langle F \rangle = \frac{1}{T} \int_0^T \! F \, \mathrm{d}t = \frac{L}{a^2\sqrt{1-e^2}},
\end{equation}
\begin{align} \label{eq:Tt}
\langle T_{eq} \rangle = \frac{1}{T} \int_0^T \! T_{eq} \, \mathrm{d}t &= T_o\left[ \frac{(1-A)L}{\beta \epsilon a^2}\right] ^\frac{1}{4}\frac{2\sqrt{1+e}}{\pi} \; \mathbf{E}\left ( \sqrt{\frac{2e}{1+e}} \right ) \\
&\approx T_o\left[ \frac{(1-A)L}{\beta \epsilon a^2}\right] ^\frac{1}{4} \left[ 1 - \tfrac{1}{16}e^2 - \tfrac{15}{1024}e^4 + O(e^6) \right],
\end{align}
where $\mathbf{E}$ is the complete elliptic integral of the second kind \citep{MathWorld, GSL} and $a$ is the semi-major axis. Equation \ref{eq:rt} is the correct physical average distance of planets, but equation \ref{eq:rs} is sometimes used instead. Equation \ref{eq:Ft} is a well known expression for the average stellar flux of elliptical orbits \citep{2002IJAsB...1...61W}. Equation \ref{eq:Tt} is a new result of this study. These equations show that the average distance ($a \leq \langle r \rangle < \frac{3}{2}a$) and stellar flux ($F|_{e=0} \leq \langle F \rangle < \infty$) increases with eccentricity while the average equilibrium temperature ($T_{eq}|_{e=0} \leq \langle T_{eq} \rangle < \frac{2\sqrt{2}}{\pi} T_{eq}|_{e=0}$) slowly decreases until converging to $\sim90\%$ of the equilibrium temperature for circular orbits (figure \ref{fig.t}).

\section{The Habitable Zone} \label{sec:hz}

The HZ can be defined either in terms of distance, stellar flux, or equilibrium temperature for circular orbits, using equations \ref{eq:F} and \ref{eq:T} to convert from one another. This is not trivial for elliptical orbits because these quantities diverge differently with eccentricity. The usual approach is to compare the average stellar flux of the planet in an elliptical orbit (equation \ref{eq:Ft}) with the stellar flux limits of the HZ, the so called `mean flux approximation' \citep{2016A&A...591A.106B}. Equivalently, \cite{2008AsBio...8..557B} suggested an HZ definition for elliptical orbits based on the mean flux approximation, but in terms of orbital distance. Their eccentric habitable zone (EHZ) is the range of orbits for which a planet receives as much flux over one orbit as a planet on a circular orbit in the HZ. This consists of comparing the semi-major axis of the planet with the HZ limits for circular orbits scaled by $(1-e^2)^{-1/4}$.

Here we prefer to avoid redefining the limits of the HZ, which adds a complication for multiplanetary systems (\emph{i.e.}, each planet having a different HZ). Instead, we suggest to define some effective distance for planets in elliptical orbits, not necessarily equal to their average orbital distance as given by equations \ref{eq:rs} or \ref{eq:rt}. Based on the mean flux approximation, the effective flux distance $r_{F}$, the equivalent circular orbit with the same average stellar flux $\langle F \rangle$ as the elliptic orbit is given by
\begin{equation} \label{eq:mfd1}
r_{F} = a(1-e^2)^{1/4}.
\end{equation}
Alternatively, using equation \ref{eq:Tt} we define a new effective thermal distance $r_{T}$, the equivalent circular orbit with the same average equilibrium temperature $\langle T_{eq} \rangle$ as the elliptic orbit, given by
\begin{align} \label{eq:mfd2}
r_{T} &= a \left[ \frac{2\sqrt{1+e}}{\pi} \; \mathbf{E}\left ( \sqrt{\frac{2e}{1+e}} \right) \right] ^ {-2} \\
&\approx a \left[ 1 + \tfrac{1}{8}e^2 + \tfrac{21}{512}e^4 + O(e^6) \right].
\end{align}

The effective distance based on the mean flux approximation (equation \ref{eq:mfd1}) and our \emph{mean thermal approximation} (equation \ref{eq:mfd2}) are not equivalent, one decreasing while the other increasing with eccentricity, respectively. The difference between these two solutions are less than 5\% for eccentricities below 0.7, so our solution is a small correction that is more significant for high eccentric orbits. Nevertheless, we argue that the mean thermal approximation provides a better characterization of the limits of the HZ than the mean flux approximation for elliptical orbits since the HZ is constrained by the thermal regime of the planet. Therefore, we suggest that equation \ref{eq:mfd2} is better to determine whether planets in elliptical orbits are within the HZ limits\footnote{VPL's HZ Calculator: \href{http://depts.washington.edu/naivpl/content/hz-calculator}{http://depts.washington.edu/naivpl/content/hz-calculator}}.


\section{Discussion} \label{sec:discussion}


Our temporal solution shows that the average equilibrium temperature of planets slightly decreases with eccentricity until converging to a value about 10\% less than for circular orbits. This result might seem contradictory because the average stellar flux strongly increases with eccentricity; however, the average distance also increases with increasing eccentricity. The net effect is a small decrease in the equilibrium temperature with eccentricity as long as the planetary albedo stays constant throughout the orbit. That might be the case for bodies without atmospheres; however, the bond albedo of planets with condensables (\emph{e.g.}, oceans) in elliptic orbits might change significantly from the hot periastron (\emph{e.g.}, due to the absorption from water vapor) to the cold apastron (\emph{e.g.}, due to the reflectivity of ice). 


We were able to reproduce well known expressions for the average distance and average stellar flux (equations \ref{eq:rs}, \ref{eq:rt}, and \ref{eq:Ft}) among our analytic solutions. The complete elliptic integral of the second kind were calculated using the GNU Scientific Library function \texttt{gsl\_sf\_ellint\_Ecomp} \citep{GSL} and the Mathematica function \texttt{EllipticE} (note that the Mathematica implementation of this function requires the square of the argument). Our results were also validated with numerical solutions. The equations \ref{eq:F} and \ref{eq:T} were independently numerically integrated for a full orbit in IDL using the \texttt{RK4} procedure, which is a fourth-order Runge-Kutta method. Both the analytic and numerical solutions agree as shown in figures \ref{fig.s} and \ref{fig.t}.

\begin{figure}
\epsscale{1.0}
\plotone{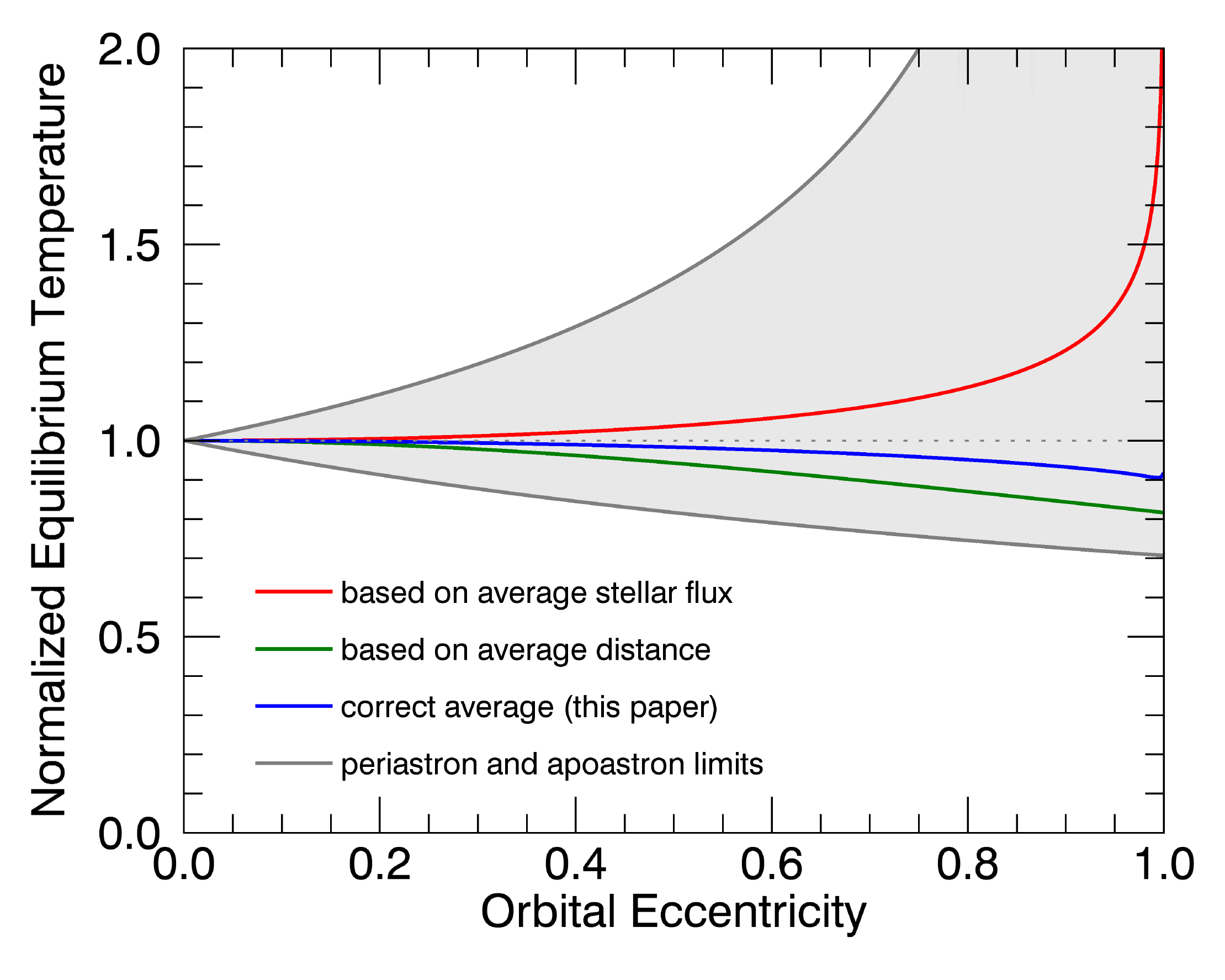}
\caption{Comparison of different averages for orbital equilibrium temperature as a function of eccentricity (normalized with respect to the circular orbits values). Averages based on stellar flux (red line) (\emph{i.e.}, the mean flux approximation) or distance (green line) are incorrect interpretations used in the literature (equations \ref{eq:Twrong1} and \ref{eq:Twrong2}). The correct average (blue line) was derived in this paper (equation \ref{eq:Tt}). All averages are within the minimum value at apastron and the maximum at periastron (shaded region between the grey lines).}
\label{fig.temp}
\end{figure}

As discussed in the introduction \S\ref{sec:intro}, there are different and incorrect interpretations in the literature of the equilibrium temperature of planets in elliptical orbits (equations \ref{eq:Twrong1} and \ref{eq:Twrong2}). They might introduce large errors especially for large eccentricities (figure \ref{fig.temp}). The most common interpretation is one based on the average stellar flux (equation \ref{eq:Twrong1}), also known as the `mean flux approximation'. This interpretation gives errors larger than 5\% for the average equilibrium temperature of planets with eccentricities higher than 0.48, and over 10\% for $e > 0.64$. In fact, the interpretation based on the average distance (equation \ref{eq:Twrong2}) is more consistent with our result (\emph{i.e.}, also decreases with eccentricity). In any case, average values should always be between the corresponding minimum orbital distance at periastron $r_p=a(1-e)$ and maximum at apastron $r_a=a(1+e)$.


A similar, but incorrect analytic solution to our average equilibrium temperature from equation \ref{eq:Tt} was proposed by \cite{2014MNRAS.440.3685B} (equation 13 in their paper). Their solution suggests that the equilibrium temperature increases instead with eccentricity, but it is not self-consistent and for eccentricities below 0.8 gives values larger than the equilibrium temperature at periastron (\emph{i.e.}, the maximum value). For example, the solution of \cite{2014MNRAS.440.3685B} gives 230 K for the average equilibrium temperature for Mars based on its orbital parameters and assuming a 0.25 bond albedo. This average value is outside the possible range of 200 K to 220 K between apastron and periastron, respectively. Our solution gives 210 K, which is consistent with the NASA Ames Mars General Circulation Model \citep{2013Icar..223..619H}. Our calculated values are also in agreement from those in the literature for other Solar System objects (table \ref{table:planet}).


Unfortunately, previous studies on the climate of planets in elliptical orbits did not include equilibrium temperature estimates to directly compare with our results. Nevertheless, our results might explain previous conflicting studies, using surface temperature as a proxy. There are different ways to study the effect of eccentricity on the climate of planets (\emph{e.g.}, at constant semi-major axis, stellar flux, etc.). \cite{2016A&A...591A.106B} studied the effect of eccentricity on the climate of tidally-locked ocean planets using a Global Climate Model LMDz under constant stellar flux. They showed that the higher the eccentricity of the planet and the higher the luminosity of the star, the less reliable is the mean flux approximation. A similar conclusion was previously obtained by \cite{2010ApJ...721.1295D} for Earth-like planets in high eccentric orbits with a one-dimensional energy balance climate model (EBM). Our equilibrium temperature estimates positively correlate with the surface temperature results of \cite{2016A&A...591A.106B}, which show that temperature decreases with eccentricity, contrary to the mean flux approximation (table \ref{table:gcm}). In this case, our results characterize the decreasing thermal response with eccentricity and corresponding increase in effective orbital distance.


Our results are more relevant for planets in very eccentric orbits $e>0.5$ and constant albedo (\emph{e.g.}, hot-jupiters). In any case, they seem better to describe the thermal regime of planets than the mean flux approximation (figure \ref{fig.temp}). Temperatures could increase with eccentricity as also shown by climate models \citep{2002IJAsB...1...61W, 2010ApJ...721.1295D}. Planets with condensable species (\emph{e.g.}, Earth-like) could experience strong variations on their albedo and heat redistribution during their elliptical orbit. \cite{1993Icar..101..108K} showed that the planetary albedo not only depends on the surface and atmospheric properties of a planet, but it is also affected by both the stellar flux and the spectrum of the star. For example, it could go below 0.2 for Earth-like planets receiving over 20\% the stellar flux that Earth receives around a Sun-like or cooler star. Heat redistribution is also more efficient in longer period tidally-locked planets \citep{2016A&A...591A.106B}. Future studies will consider these important dynamical corrections.

\section{Conclusion} \label{sec:conclusion}

Here we determined analytic solutions of the spatial and temporal averages of orbital distance, stellar flux, and equilibrium temperature for planets in elliptical orbits and constant albedo. The solutions were validated (see \S\ref{sec:discussion}) by reproducing  known solutions with the same approach, obtaining equal results with numerical solutions, comparing with some Solar System bodies, and producing results consistent with previous climate models for planets in elliptic orbits. In particular, we determined that:
\begin{enumerate}
\item
The average equilibrium temperature $\langle T_{eq} \rangle$ of planets in elliptical orbits slowly decreases with eccentricity $e$ until a converging value $\sim90\%$ ($\frac{2\sqrt{2}}{\pi}$) the equilibrium temperature for circular orbits, assuming a constant albedo $A$ and heat redistribution $\beta$. The temperature is given by
\begin{equation} \label{eq:Tt2}
\langle T_{eq} \rangle=T_o\left[ \frac{(1-A)L}{\beta \epsilon a^2} \right] ^\frac{1}{4}\frac{2\sqrt{1+e}}{\pi} \; \mathbf{E}\left ( \sqrt{\frac{2e}{1+e}} \right),
\end{equation}
where $T_o$= 278.5 K, $L$ is the star luminosity (solar units), $a$ is the semi-major axis (AU), $\epsilon$ is the emissivity, and $\mathbf{E}$ is the complete elliptic integral of the second kind \citep{MathWorld, GSL}.
\item
The potential stability of surface liquid water on rocky planets in elliptical orbits is better characterized by a mean thermal approximation rather than the mean flux approximation. Using observational data, a planet is inside the HZ if their effective thermal distance $r_{T}$, given by
\begin{equation} \label{eq:mfd3}
r_{T} = a \left[ \frac{2\sqrt{1+e}}{\pi} \; \mathbf{E}\left ( \sqrt{\frac{2e}{1+e}} \right) \right] ^ {-2},
\end{equation}
is within the HZ limits for circular orbits as defined by \cite{1993Icar..101..108K}. In general, planets move farther outside the HZ (\emph{i.e.}, become colder) with increasing eccentricity if all other conditions stay constant; however, orbital variations of albedo, greenhouse, and heat redistribution might accentuate or invert this trend.
\item
All climate models of planets should include at least global calculations of albedo $A$, normalized greenhouse $g$, equilibrium $T_{eq}$, and surface temperatures $T_s$ as a function of orbital time, not just orbital longitude. The distinction between the use of the surface temperature or the effective surface temperature should also be clear \citep{2013Icar..223..619H}. These global parameters help to compare different climate models and to translate them to simple parametric functions that are necessary to determine whether planets are in the HZ from observational data.
\end{enumerate}

The results of this paper might be used to reevaluate the effective location in the HZ of many exoplanets with known eccentricity (using equation \ref{eq:mfd3}), especially those small enough to be considered potentially habitable. For example, the planet Proxima Cen b is probably in a circular orbit \citep{2016Natur.536..437A}, but the outer two planets in the red-dwarf star Wolf 1061 are in eccentric orbits crossing the HZ \citep{2016ApJ...817L..20W, 2016arXiv161209324K}. Also, many giant planets with elliptical orbits that cross the HZ might support habitable exomoons \citep{2013MNRAS.432.2994F, 2013ApJ...774...27H}. Results for individual exoplanets are available and updated in the \emph{PHL's Exoplanet Orbital Catalog}\footnote{Exoplanet Orbital Catalog: \href{http://phl.upr.edu/projects/habitable-exoplanets-catalog/catalog}{http://phl.upr.edu/projects/habitable-exoplanets-catalog/catalog}}.


\acknowledgments

This work was supported by the Planetary Habitability Laboratory (PHL) and the Center for Research and Creative Endeavors (CIC) of the University of Puerto Rico at Arecibo (UPR Arecibo). Part of this study was originally presented in a poster at the 2014 STScI Spring Symposium: \emph{Habitable Worlds Across Space And Time} (April 28 - May 1, 2014). This research has made use of the NASA Exoplanet Archive, which is operated by the California Institute of Technology, under contract with the National Aeronautics and Space Administration under the Exoplanet Exploration Program. We gratefully acknowledge the anonymous referee, whose comments improved the quality of the paper.



\clearpage

\begin{table}
\begin{threeparttable}
\caption{Calculated temporal averages of orbital distance $\langle r \rangle$, stellar flux $\langle F \rangle$, and equilibrium temperature $\langle T_{eq} \rangle$ for some Solar System bodies.}
\bigskip
\begin{tabular}{ l c c c c | c c c }
 \hline
 \hline
 & \multicolumn{4}{c}{Literature Values\tnote{a}} & \multicolumn{3}{c}{Temporal Averages\tnote{b}} \\
 \hline
 Name & $a$ (AU) & $e$ & $A$ & $T_{eq}$ (K) &
 	$\langle r \rangle$ (AU) & $\langle F \rangle$ & $\langle T_{eq} \rangle$ (K) \\ 
 \hline
 Mercury & 0.38709927  & 0.20563593 & 0.068 & 439.6 & 0.395  & 6.819 & 439 \\  
 Venus   & 0.72333566  & 0.00677672 & 0.77  & 226.6 & 0.723  & 1.911 & 227 \\
 Earth   & 1.00000261  & 0.01671123 & 0.306 & 254.0 & 1.000  & 1.000 & 254 \\
 Moon    & 1.00000261  & 0.01671123 & 0.11  & 270.4 & 1.000  & 1.000 & 271 \\
 Mars    & 1.52371034  & 0.0933941  & 0.250  & 209.8 & 1.530  & 0.433 & 210 \\
 Pluto   & 39.48211675 & 0.2488273  & 0.4   & 37.5  & 40.704 & 0.001 & 39 \\
 \hline
 \hline
\end{tabular}
\label{table:planet}
	\begin{tablenotes}
	\small
\item[a]{Literature values are from the orbital elements data of NASA Solar System Dynamic: \href{http://ssd.jpl.nasa.gov/?planets}{http://ssd.jpl.nasa.gov/?planets}, and albedo and equilibrium temperature data of the NASA Planetary Fact Sheets: \href{http://nssdc.gsfc.nasa.gov/planetary/planetfact.html}{http://nssdc.gsfc.nasa.gov/planetary/planetfact.html}.}
\item[b]{Average values calculated from the literature values and the temporal average equations \ref{eq:rt}, \ref{eq:Ft}, and \ref{eq:Tt}, respectively.}
	\end{tablenotes}
\end{threeparttable}

\end{table}

\clearpage



\begin{table}
\begin{threeparttable}
\caption{Estimated average equilibrium temperature $\langle T_{eq} \rangle$ and effective thermal distance $r_T$ for a modeled ocean planet with increasing eccentricity, but under a constant average stellar flux equal to Earth.}
\bigskip
\begin{tabular}{ c c c c | c c }
 \hline
 \hline
 \multicolumn{4}{c}{Planet Model\tnote{a}} & \multicolumn{2}{c}{Estimated Values} \\
 \hline
 $e$ & $a$ (AU) & $A$ & $\langle T_s \rangle$ (K) &
 $\langle T_{eq} \rangle$\tnote{b} (K) &
 $r_T$\tnote{c} (AU) \\
 \hline
 0.0 & 1.000 & 0.250 & 267 & 259 & 1.000 \\
 0.2 & 1.003 & 0.295 & 263 & 253 & 1.008 \\
 0.4 & 1.045 & 0.350 & 255 & 242 & 1.067 \\
 \hline
 \hline
\end{tabular}
\label{table:gcm}
	\begin{tablenotes}
	\small
	\item[a]{Global Climate Model LMDz for a tidally-locked ocean planet in an elliptical orbit around a Sun-like star with $L=1$ and $\langle F \rangle=1$ \citep{2016A&A...591A.106B}.}
	\item[b]{Average equilibrium temperature from equation \ref{eq:Tt} with $\epsilon=1$ and $\beta=1$.}
	\item[c]{Effective thermal distance from equation \ref{eq:mfd2}.}
	\end{tablenotes}
\end{threeparttable}

\end{table}

\clearpage


\clearpage

\clearpage

\end{document}